# Activation Volume in the Density Scaling Regime: Equation of State and Its Test by Using Experimental and Simulation Data

A. Grzybowski,* K. Koperwas, A. Swiety-Pospiech, K. Grzybowska, and M. Paluch

*Institute of Physics, University of Silesia, Uniwersytecka 4, 40-007 Katowice, Poland*

In this paper, a formalism for the activation volume of glass forming materials is suggested. An isothermal equation of state for the activation volume is formulated, which is extended to a generalized equation of state that describes the activation volume as a function of temperature and pressure. Both the equations of state are very successfully validated by using experimental and simulation data collected for supercooled Kob-Andersen binary Lennard-Jones liquid and materials from various material groups such as van der Waals liquids, polymers, protic ionic liquids, and strongly hydrogen bonded liquids. Some predictions based on these equations of state for the activation volume are also very satisfactorily verified in case of each considered system, especially a kind of the activation volume scaling with the scaling exponent that also constitutes the slope of the expected linear pressure dependence of the isothermal bulk modulus for the activation volume is confirmed. The until recently unexpected negative value of the slope are explained in case of the systems that obey the thermodynamic scaling law at least to a good approximation.

* email: *andrzej.grzybowski@us.edu.pl*

## I. Introduction

In the last two decades, the glass transition physics has been considerably enriched with many new results obtained from high pressure experiments [1 - 13], which have extended the traditional study of the glass transition in the temperature domain by the pressure effects and the corresponding influence of density changes, where the latter is possible to determine if we know temperature-pressure volumetric data usually described by an appropriate equation of state (EOS). The high pressure techniques enable to explore physical phenomena observed for systems which are approaching the glass transition along various thermodynamic paths. Besides a typical method for a glass formation, which is the vitrification of a liquid by sufficiently fast isobaric cooling of the liquid both at ambient and elevated pressures, another method highly exploited for the glass formation is the isothermal compression of a liquid. These two kinds of experiments provide in the natural way two kinds of information [1,2] on the examined material, i.e., its sensitivity to temperature and pressure changes near the glass transition, which can be quantified respectively by the isobaric fragility parameter, $m_p = \partial \log_{10}(\tau)/\partial(T_g/T)|_p$, calculated usually at the glass transition temperature $T_g$, and the activation volume originally derived by Williams [14] within the framework of the Eyring transition state theory at isothermal conditions as follows

$$\upsilon_{act} = RT \frac{\partial \ln \tau}{\partial p}\bigg|_T \qquad (1)$$

where the structural relaxation time $\tau$ has been considered, because this dynamic quantity is used in all tests reported in this paper, and $R$ is the gas constant.

It should be noted that the parameters $m_p$ and $\upsilon_{act}$ are not independent, because there is a relation between them [15]. Thus, the mentioned distinguishing between $m_p$ and $\upsilon_{act}$ should not be treated restrictively. For instance, it is commonly known that the isobaric fragility depends on pressure, i.e., usually decreases with increasing pressure [1,2]. As a result, the parameter $m_p$ also provides some information on the sensitivity of materials to pressure changes. However, the activation volume is the basic parameter that characterizes how pressure changes affect molecular dynamics of a given material near the glass transition. This feature of $\upsilon_{act}$ demands that we discuss the activation volume in the context of density scaling of molecular dynamics near the glass transition, which is one of the most appealing ideas proposed in recent years to describe properties of glass formers at elevated pressure [1].

The density scaling of molecular dynamics of viscous systems has been intensively explored in the last decade to gain a better insight into the glass transition and related phenomena. These promising investigations have been started with phenomenological observations [16 - 23] which have resulted in plotting isothermal and isobaric experimental dependences of dynamic quantities such as structural relaxation time $\tau$ and viscosity $\eta$ onto one master curve as a function of the scaling quantity, $\Gamma = \rho^\gamma/T$, where $\rho$ – density, $T$ – temperature, and the scaling exponent $\gamma$ is a material constant. A lot of effort has been later put into understanding this scaling phenomenon and finding its theoretical explanation. The most common point of view on the theoretical grounds for the power-law density scaling also called thermodynamic scaling is based on a dominant role of short-range interactions in viscous systems, which are characterized by small intermolecular distances. In a consequence, it is usually suggested [24 - 31] that the thermodynamic scaling can be explained by means of the short range effective approximation, $U_{eff}(r) = 4\varepsilon(\sigma/r)^{m_{IPL}} - A_t$, of the generalized

Lennard-Jones (LJ) potential [32], $U_{LJ}(r) = 4\varepsilon\left[(\sigma/r)^m - (\sigma/r)^n\right]$, where $A_t$ is some constant (or linear) small attractive background and the inverse power-law exponent, $m_{IPL} = 3\gamma_{IPL} > m$, is straightforwardly related to the exponent $\gamma_{IPL}$, which can be identified with the scaling exponent γ that enables to scale structural relaxation times and viscosities.

Moreover, it has been shown by using theoretical and simulation studies [26 - 28] that each system which obey the power-law density scaling should fulfill a strong linear correlation between the system virial $W$ and the system potential energy $U$: $\Delta W(t) \cong \gamma_{WU}\Delta U(t)$, where $\Delta U(t) = U(t) - <U>$ and $\Delta W(t) = W(t) - <W>$ are respectively deviations of the instantaneous values $U(t)$ and $W(t)$ from their thermal averages $<U>$ and $<W>$. It has been also argued that the $WU$ correlation coefficient $\gamma_{WU}$ can be identified with the exponent $\gamma_{IPL}$ if the short range effective potential $U_{eff}$ is assumed to be relevant to the density scaling of molecular dynamics. This theoretical argumentation has been confirmed by results of molecular dynamics (MD) simulations with simple force fields based on the Lennard-Jones potential [26 - 28]. These simulation studies have also enabled to establish a good agreement between the $WU$ correlation coefficient $\gamma_{WU}$ and the exponent γ that scales dynamic quantities such as structural relaxation times, viscosities, and diffusivities. Recently, Pedersen et al. [33] have shown that the isochoric heat capacity calculated from simulations of the Kob-Andersen binary Lennard-Jones (KABLJ) liquid [34] can be also scaled versus the quantity $\rho^\gamma/T$ with the same value of γ as that used to scale dynamic quantities. However, such power-law density scaling of volumetric data has been found to be impossible in the exact sense [19,25,31,35 - 40].

To explore the problem with the scaling of pressure-volume-temperature (PVT) data, we have recently formulated two equations of state, which originate from different grounds [38,39], i.e., the same short range effective $U_{eff}$ that underlies the WU correlation and a modified definition of the isothermal bulk modulus, where the exponent $\gamma_{EOS}$ is introduced only in the following mathematical way, $B_T = -\gamma_{EOS}\left(\partial p / \partial \ln \upsilon^{\gamma_{EOS}}\right)_T$. The latter approach yields the equation of state, which is simpler to use. Thus, we exploit it herein in the following form,

$$\left(\frac{\upsilon(T,p_0)}{\upsilon(T,p)}\right)^{\gamma_{EOS}} = 1 + \frac{\gamma_{EOS}}{B_T(p_0)}(p-p_0) \qquad (2)$$

where the temperature dependence specific volume $\upsilon(T,p_0)$ at a reference pressure $p_0$ is typically parametrized by the quadratic temperature function, $\upsilon_0(T) = \upsilon(T,p_0) = A_0 + A_1(T-T_0) + A_2(T-T_0)^2$, and the temperature dependence of the isothermal bulk modulus at $p_0$ by the exponential temperature function $B_T(p_0) = B_{T_0}(p_0)\exp(-b_2(T-T_0))$, where the fitting parameter $b_2$ is approximately constant independently of the choice of $T_0$ in the supercooled region, and the fixed reference state $(T_0, p_0)$ can be usually defined at the glass transition temperature and ambient pressure.

It is worth noting that the shared assumption exploited to derive both the mentioned equations of state relies on a small compressibility, which characterizes materials near the glass transition and allows to limit their range of validity to the case of the linear pressure dependence of the isothermal bulk modulus,

$$B_T(p_0) = B_{T_0}(p_0) + \gamma_{EOS}(p-p_0). \quad (3)$$

Tests of these equations of state by using experimental and simulation data have shown that both these equations yield numerically very close values of $\gamma_{EOS}$ for a given material, which can be used to a kind of scaling of volumetric data. Since the exponent $\gamma_{EOS}$ can be straightforwardly related to the exponent $\gamma_{IPL}$ from the short range effective potential $U_{eff}$ according to the assumption made to derive one of these equations of state, the exponent $\gamma_{EOS}$ should correspond to both the *WU* correlation coefficient $\gamma_{WU}$ and the exponent γ that scales dynamic quantities. In fact, we have very recently established [40] such a correspondence $\gamma_{EOS} \approx \gamma_{WU} \approx \gamma$ for simulation data obtained within the framework of the KABLJ model. However, this relation is broken in case of experimental data, for which we have observed a considerable discrepancy $\gamma_{EOS} \gg \gamma$ [36-39].

In the context of the hot debate about the scaling phenomena near the glass transition, it is important to answer the question *whether the activation volume can be scaled or not* and to establish consequences of the answer. This paper is devoted to this issue. We show herein how to describe the activation volume in the density scaling regime. We find which kind of scaling is valid for the activation volume and we define the proper scaling exponent for this quantity. What is more, we formulate an equation of state for the activation volume. This EOS leads us to very interesting predictions about properties of molecular dynamics near the glass transitions, which are verified by using experimental and simulation data.

## II. Activation Volume in the Density Scaling Regime

According to the general approach to the density scaling of dynamic quantities [25,41] the structural relaxation time $\tau$ of viscous systems can be described by some general density-temperature function

$$\ln \tau(\rho,T) = f\left(\frac{h(\rho)}{T}\right) \qquad (4)$$

where $h(\rho)$ is only a density function.

Then, the activation volume can be derived from Eqs. (1) and (4) as follows

$$\upsilon_{act} = RT \frac{f'(x)}{T} \frac{\partial h(\rho)}{\partial \rho} \frac{\partial \rho(T,p)}{\partial p}\bigg|_T \qquad (5)$$

where $f'(x)$ is a partial derivative of the external function with respect to $x = h(\rho)/T$.

Since $\dfrac{\partial h(\rho)}{\partial \rho} = h(\rho)\dfrac{\partial \ln h(\rho)}{\partial \rho}$ and $\dfrac{\partial \rho(T,p)}{\partial p}\bigg|_T = \dfrac{\rho}{B_T(p)}$, where

$$B_T(p) \equiv -\upsilon \frac{\partial p}{\partial \upsilon}\bigg|_T \qquad (6)$$

is the isothermal bulk modulus of the specific volume $\upsilon$, the activation volume

$$\upsilon_{act} = R \frac{f'(x)}{B_T(p)} h(\rho) \frac{\partial \ln h(\rho)}{\partial \ln \rho} \qquad (7)$$

Then, the inverse reduced activation volume in the density scaling regime can be expressed as follows

$$\frac{\upsilon_{act}(T,p_0)}{\upsilon_{act}(T,p)} = \frac{B_T(p)}{B_T(p_0)} \frac{f'(x_0)}{f'(x)} \frac{h(\rho(T,p_0))}{h(\rho(T,p))} \qquad (8)$$

where $x = h(\rho(T,p))/T$ and $x_0 = h(\rho(T,p_0))/T$.

According to the density scaling [41,42], in general, the scaling exponent $\gamma \equiv \dfrac{\partial \ln h(\rho)}{\partial \ln \rho}$. It is also valid in the case of the power-law density scaling called also thermodynamic scaling, because then $h(\rho) = \rho^\gamma$. To find an explicit form of Eq. (8) we need to assume an expression for the function $f$ in Eq. (4). Until recently, the best model to describe the thermodynamic scaling of dynamic quantities such as structural relaxation times of viscosities has been considered [1] the density-temperature version [43] of the Avramov entropic model [44],

$$\ln \tau(\rho, T) = \ln \tau_0 + \left(\frac{A\rho^\gamma}{T}\right)^D \tag{9}$$

Since $\ln\tau_0$, A, D and $\gamma$ are fitting parameters in Eq. (9), and the density scaling function can be considered as $h(\rho) = \rho^\gamma$, the derivative $f'(x) = DA^D x^{D-1}$ where $x = \rho^\gamma / T$. In this way, the activation volume in the case of the power-law density scaling can be expressed as follows

$$\upsilon_{act} = RT \frac{\gamma D}{B_T(p)} \left(\frac{A\rho^\gamma}{T}\right)^D \tag{10}$$

and then Eq. (8) with $\rho = \upsilon^{-1}$ provides us an explicit expression for the inverse reduced activation volume, $\frac{\upsilon_{act}(T, p_0)}{\upsilon_{act}(T, p)} = \frac{B_T(p)}{B_T(p_0)} \left(\frac{\upsilon(T, p)}{\upsilon(T, p_0)}\right)^{\gamma D}$. From the latter equation and Eq. (2), one can find the following important relations

$$\frac{\upsilon_{act}(T, p_0)}{\upsilon_{act}(T, p)} = \left(\frac{\upsilon(T, p_0)}{\upsilon(T, p)}\right)^{\gamma_{EOS} - \gamma D} \tag{11}$$

$$\frac{\upsilon_{act}(T, p_0)}{\upsilon_{act}(T, p)} = \left(\frac{B_T(p_0)}{B_T(p)}\right)^{-(\gamma_{EOS} - \gamma D)/\gamma_{EOS}} \tag{12}$$

which show that the inverse reduced activation volume is some power functions of the inverse reduced volume and the inverse reduced isothermal bulk modulus.

## III. Equation of State for the Activation Volume

### (a) Isothermal EOS for $\upsilon_{act}$

The relations given by Eqs. (11) and (12) suggest that the activation volume can obey an analogous isothermal EOS to Eq. (2). To derive this EOS for the activation volume we can apply the method previously used [38] to find the isothermal version of Eq. (2).

It should be noted that the isothermal bulk modulus for the activation volume can be defined as

$$B_{act} \equiv -\upsilon_{act} \left.\frac{\partial p}{\partial \upsilon_{act}}\right|_T \tag{13}$$

by analogy with the definition of the isothermal bulk modulus for volume and we can exploit the same mathematical trick $B_{act} = -\gamma_{act} \left(\partial p / \partial \ln \upsilon_{act}^{\gamma_{act}}\right)_T$ as that used to find the EOS for the specific volume given by Eq. (2). Then, we can consider the modified definition of $B_{act}$ as a differential equation. A general solution of this equation, $\upsilon_{act}^{-\gamma_{EOS}} = \exp\left(\gamma_{act} \int \frac{dp}{B_{act}(p)}\right)$, leads with an initial condition $\upsilon_{act0} = \upsilon_{act}(T, p_0)$ to its particular solution $(\upsilon_{act0}/\upsilon_{act})^{\gamma_{act}} = \exp\left(-\gamma_{act} \int \frac{dp}{B_{act}(p)}\right)\bigg|_{p=p_0} \exp\left(\gamma_{act} \int \frac{dp}{B_{act}(p)}\right)$. Then, a first-order Taylor series expansion of this particular solution about $p=p_0$ yields the isothermal EOS for the activation volume

$$\left(\frac{\upsilon_{act}(T, p_0)}{\upsilon_{act}(T, p)}\right)^{\gamma_{act}} = 1 + \frac{\gamma_{act}}{B_{act}(T, p_0)}(p - p_0) \tag{14}$$

where $B_{act}(T, p_0)$ is the isothermal bulk modulus for the activation volume at the reference state defined by $p_0$ at a constant temperature $T$.

It has been already mentioned that Eq. (2) satisfies the linear pressure dependence of the isothermal bulk modulus given by Eq. (3). Similarly, one can easily derive from the definition of $B_{act}$ by differentiating the EOS (Eq. (14)) for $\upsilon_{act}$ that it implies the following linear pressure dependence of the isothermal bulk modulus for the activation volume,

$$B_{act}(T, p) = B_{act}(T, p_0) + \gamma_{act}(p - p_0) \tag{15}$$

The above equation shows the basic physical meaning of the arbitrarily introduced parameter of the exponent $\gamma_{act}$, which is defined by the derivative $\gamma_{act} = \partial B_{act}(T, p)/\partial p\big|_{p=p_0}$ at a given $T$. It means that Eq. (14) is valid if the pressure dependence of the isothermal bulk modulus for the activation volume is linear and its slope is independent of thermodynamic conditions. It is worth noting that we should be able to predict values of the parameters $\gamma_{act}$ and $B_{act}(T, p_0)$ without any need to determine the activation volume values. Taking into account the EOS for the specific volume (Eq. (2)) and the relation given by Eq. (11), we can find an auxiliary equation

$$\left(\frac{\upsilon_{act}(T, p_0)}{\upsilon_{act}(T, p)}\right)^{\frac{\gamma_{EOS}}{\gamma_{EOS} - \gamma D}} = 1 + \frac{\gamma_{EOS}}{B_T(p_0)}(p - p_0), \tag{16}$$

Then, calculating the isothermal bulk modulus for the activation volume $B_{act}(T, p)$ from Eq. (16) and comparing the obtained equation, $B_{act}(T, p) = \frac{B_T(p_0)}{\gamma_{EOS} - \gamma D} + \frac{\gamma_{EOS}}{\gamma_{EOS} - \gamma D}(p - p_0)$, with Eq. (15), we find the following dependences

$$\gamma_{act} = \frac{\gamma_{EOS}}{\gamma_{EOS} - \gamma D} \tag{17}$$

$$B_{act}(T, p_0) = \frac{B_T(p_0)}{\gamma_{EOS} - \gamma D} \tag{18}$$

Thus, the relations given by Eqs. (11) and (12) can be expressed also by replacing their exponents with $\gamma_{EOS}/\gamma_{act}$ and $-1/\gamma_{act}$, respectively. Moreover, if we take into account the pressure dependences given by Eqs. (3) and (15) or the definitions of $B_T$ and $B_{act}$, i.e., Eqs. (6) and (13), the isothermal equations of state for the specific and activation volumes (Eqs. (2) and (14)) imply that the scaled inverse reduced volume corresponds to the reduced bulk modulus. Such a relation is valid for both the non-activation and activation quantities. It means that there is the following unity condition

$$\frac{B_T(p_0)}{B_T(p)} \left( \frac{\upsilon(T, p_0)}{\upsilon(T, p)} \right)^{\gamma_{EOS}} = \frac{B_{act}(T, p_0)}{B_{act}(T, p)} \left( \frac{\upsilon_{act}(T, p_0)}{\upsilon_{act}(T, p)} \right)^{\gamma_{act}} = 1 \tag{19}$$

It should be noted that Eq. (14) can be also found from the definition of the isothermal bulk modulus for the activation volume by using a particular solution of Eq. (13), i.e., $\upsilon_{act0}/\upsilon_{act} = \exp\left(-\int \frac{dp}{B_{act}(p)}\right)\bigg|_{p=p_0} \exp\left(\int \frac{dp}{B_{act}(p)}\right)$, that is obtained with the initial condition $\upsilon_{act0} = \upsilon_{act}(T, p_0)$. A second-order Taylor series expansion of this particular solution about $p=p_0$, i.e., $\upsilon_{act0}/\upsilon_{act} \cong 1 + B_{act}^{-1}(T, p_0)(p - p_0) + (1/2)B_{act}^{-2}(T, p_0)[1 - \partial B_{act}(T, p)/\partial p\big|_{p=p_0}](p - p_0)^2$, and then a subsequent second-order Taylor series expansion of this second-order approximation of $(\upsilon_{act0}/\upsilon_{act})$ raised to the power $\gamma_{act}$ also lead to Eq. (14) due to vanishing its second-order term if one assumes $\partial B_{act}(T, p)/\partial p\big|_{p=p_0} = const$ for a given material in the vicinity of the glass transition and denotes it by $\gamma_{act}$. This last assumption corresponds to the mentioned linear pressure dependence of the isothermal bulk modulus for the activation volume (Eq. (15)), which is satisfied by Eq. (14). It means that we have two potentially alternative approaches: (i) the representation $(\upsilon_{act0}/\upsilon_{act})^{\gamma_{act}}$ given by Eq. (14) and based on the first-order Taylor series expansion of the solution of the differential equation coming from the definition of the isothermal bulk modulus for the activation volume modified by introducing the exponent $\gamma_{act}$, and (ii) the representation $(\upsilon_{act0}/\upsilon_{act})$ that follows directly from the definition of the isothermal bulk modulus for the activation volume given by Eq. (13) and could be used in the form approximated by the second-order Taylor series expansion presented above without reducing it to Eq. (14) by the next second-order expansion. Some reason for choosing the approach (i) is the expression used in Eq. (14), which is simpler than the second-order expansion of $(\upsilon_{act0}/\upsilon_{act})$ and expected to describe pressure dependences of the activation volume that are in general nonlinear. However, there are more essential causes to prefer Eq. (14). Its form is very similar to Eq. (2) that leads to a scaling of volumetric data with the scaling exponent $\gamma_{EOS}$ [38]. Thus, one can expect that the activation volume can be scaled in terms of Eq. (14) with the scaling exponent $\gamma_{act}$. Moreover, the found relation (Eq. (17)) between $\gamma$, $\gamma_{EOS}$, and $\gamma_{act}$ shows that the scaling of the activation volume reflects a combined effect of the density scaling of molecular dynamics near the glass transition (with the scaling exponent $\gamma$) and the scaling of volumetric data (with the scaling exponent $\gamma_{EOS}$). In other words, the proposed formalism for the activation volume predicts that the scaling of dynamics and the scaling of PVT data result in the scaling of the activation volume of glass forming materials.

*(b) Generalized EOS for $\upsilon_{act}$*

An interesting question arises whether it is possible to determine a general EOS for the activation volume. It means whether we are able to find appropriate parametrizations of the temperature-dependent parameters of Eq. (14), which are $\upsilon_{act}(T, p_0)$ and $B_{act}(T, p_0)$.

Taking into account the morphological correspondence between Eq. (14) and Eq. (2), a natural way of the generalization about the isothermal EOS for the activation volume, which suggests itself, is to attempt at applying the same parametrizations to the temperature-dependent parameters of Eq. (14) as those used to generalize [39] the originally isothermal EOS given by Eq. (2). Thus, we postulate the following temperature dependences

$$\upsilon_{act}(T, p_0) = F_0 + F_1(T-T_0) + F_2(T-T_0)^2 \tag{20}$$

$$B_{act}(T, p_0) = B_{act}(T_0, p_0)\exp(-g_2(T-T_0)) \tag{21}$$

where $F_0 = \upsilon_{aact}(p_0, T_0)$, $F_1 = \partial \upsilon_{act}(p_0, T)/\partial T \big|_{T=T_0}$, $F_2 = (1/2)\partial^2 \upsilon_{act}(p_0, T)/\partial T^2 \big|_{T=T_0}$, and $g_2 = -\partial \ln B_{act}(p_0, T)/\partial T \big|_{T=T_0}$. The latter parameter is assumed to be approximately constant and independent of the choice of $T_0$ in the considered temperature region. The value of $T_0$ can be usually fixed as the glass transition temperature at $p_0$, i.e., $T_0 = T_g(p_0)$. As a result, we find the generalized EOS for the activation volume,

$$\upsilon_{act}(T,p) = \frac{F_0 + F_1(T-T_0) + F_2(T-T_0)^2}{\left[1 + (p-p_0)g_1 \exp(g_2(T-T_0))\right]^{1/\gamma_{act}}} \quad (22)$$

where $g_1 = \gamma_{act}/B_{act}(T_0, p_0)$, and in general the reference state $(T_0, p_0)$ for Eq. (22) is chosen similarly to that for the EOS for the specific volume (Eq. (2)) [39], i.e., in order to indicate the applicability limits for Eq. (22). It means that this EOS for the activation volume with determined values of its parameters for a given material in the supercooled state should be used only to this state, and any attempts at applying this EOS, for instance, to the glassy state require establishing another set of fitted values of the EOS parameters.

**IV. Experimental and Simulation Tests and Their Discussion**

It is worth noting that the EOS given by Eq. (14) as well as their supplementations with Eqs. (20) and (21) are derived without any assumption that the power-law density scaling is valid. Thus, we test this EOS by using experimental data for representatives of different material classes, including materials for which the density scaling has been confirmed and the others. We consider herein the earlier reported dielectric and volumetric data for materials that obey the density scaling [2,39,45] such as 1,1'-bis (p-methoxyphenyl) cyclohexane (BMPC) [46,47] and phenylphthalein-dimethylether (PDE) [48,49], which belong to supercooled van der Waals liquids, and a polymer melt (PVAc) [50,51]. We also exploit experimental data for a protic ionic liquid verapamil hydrochloride (VH) [52,53], the relaxation times of which can be scaled vs $\rho^\gamma/T$ to a good approximation, although our previous thorough analyses showed that the density scaling is not perfect for VH [52]. Moreover, we use experimental dielectric and PVT data for a strongly hydrogen-bonded liquid DPG [54,55,56], for which we have confirmed that the density scaling is broken [57]. Besides the use of measurement data, the EOS for the activation volume is verified by exploiting MD simulation data obtained in the KABLJ model [40], for which one can find an effective value of the scaling exponent $\gamma$ for the structural relaxation times, although they cannot be scaled vs $\rho^\gamma/T$ in the exact sense.

*(a) Experimental Test of the Isothermal EOS for $\upsilon_{act}$ and the Scaling of $\upsilon_{act}$*

We begin the verification procedure with the isothermal EOS for the activation volume given by Eq. (14). We determine the activation volume of BMPC by using isothermal structural relaxation times established from high pressure broadband dielectric measurements [46] and fitted herein to Eq. (9) together with the isobaric structural relaxation times at ambient pressure also reported in [46]. Exploiting Eq. (2) with the earlier reported values of the EOS parameters [58], we can find a temperature-pressure function $\tau(\rho(T,p),T)$ for the fit to Eq. (9). The relaxation times and the fitted curves are shown in Fig. 1(a) as pressure functions. Then, we evaluate the pressure dependences of the activation volume for each isotherm from the definition given by Eq. (1). As can be seen in Fig. 1(b), we can successfully fit the pressure dependences $\upsilon_{act}(p)$ to Eq. (14) at the fixed reference pressure $p_0=0.1$MPa. By using the values of the parameters found from the fitting procedure to Eq. (14), we can construct a linear scaling plot for the activation volume (Fig. 1(c)), which is analogous to the PVT scaling plots earlier reported by us for the specific volume [38] analyzed in terms of the isothermal version of Eq. (2). The similarity in the scaling procedures one can predict from Eqs. (2) and (14) follows from the morphological correspondence of these EOS. However, we can observe that the scaling procedures in terms of Eqs. (2) and (14) yield the considerably different values of the scaling exponents, $\gamma \approx 7.8$ [45] and $\gamma_{act} \approx -3.45$ for BMPC. It can be explained by Eq. (17) and will be discussed later.

*(b) Experimental Tests of the Generalized EOS for $\upsilon_{act}$ and the Scaling of $\upsilon_{act}$*

The promising successful result of the test of the isothermal version of Eq. (14) by using experimental data for BMPC encourages us to verify the assumed temperature parametrizations of Eq. (14) by Eqs. (20) and (21), which lead to the generalized EOS for the activation volume (Eq. (22)). To validate Eq. (21) we find (see Fig. 2(a)) the temperature dependences of the isothermal bulk modulus $B_{act}(T)$ for the activation volume of BMPC at a few pressures, p=0.1, 20, and 40MPa, exploiting the activation volume data shown in Fig. 1(b). Since the values of the isothermal bulk modulus for the activation volumes calculated from the definition given by Eq. (13) are negative, because the activation volume increases with increasing pressure in isothermal conditions (e.g. see Fig. 1(b)), we plot appropriate logarithmic representations of the dependences $B_{act}(T)$ expressed by $\ln(-B_{act}(T))$. As a result we find that these logarithmic functions of temperature are linear to a very good approximation (Fig. 2(b)). Thus, we can assume that the

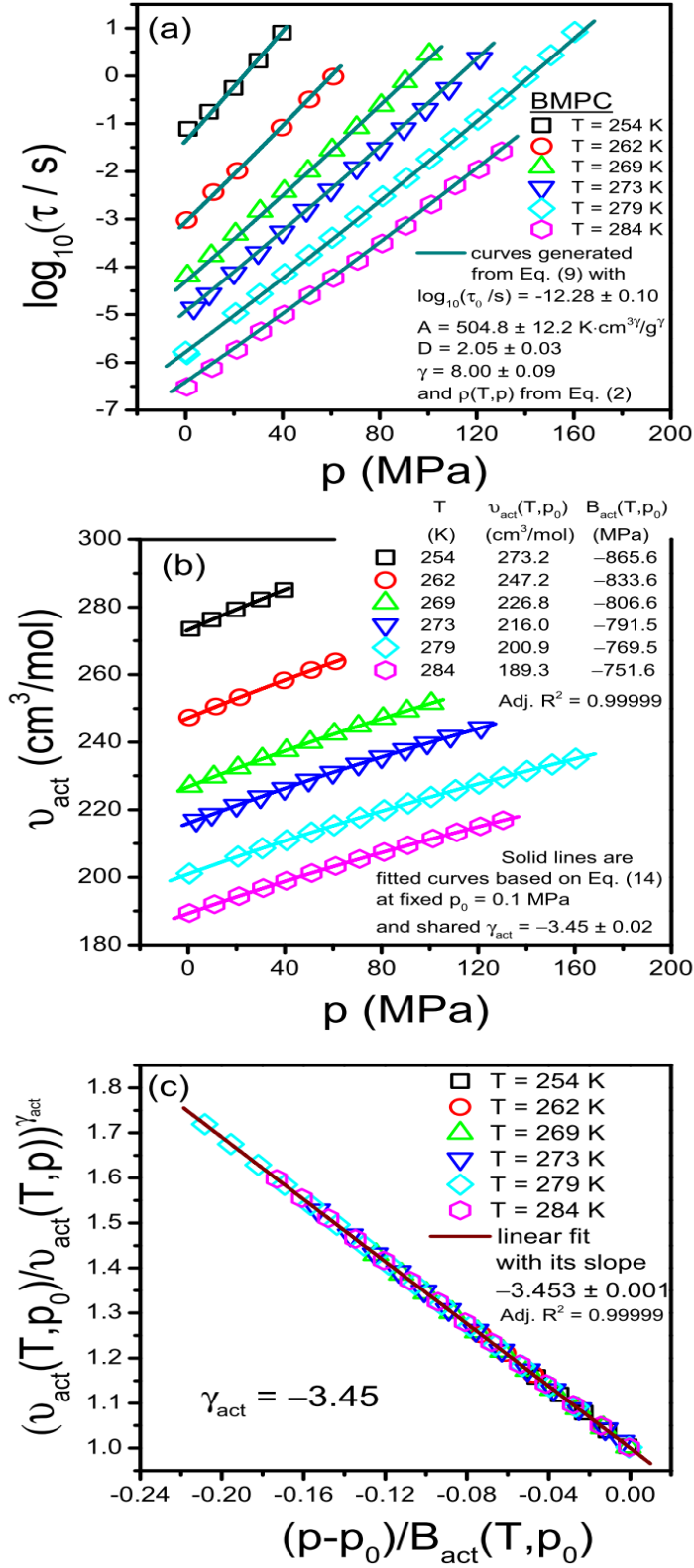

Fig. 1 (a) Plot of the pressure dependences of isothermal structural relaxation times τ of BMPC. Solid lines denote the curves generated from the shared fit of $\tau(T,\rho)$ to Eq. (9) and the temperature-pressure dependence of density established from Eq. (2) with the values of its parameters taken from [58] (b) Plot of the pressure dependences of the activation volume of BMPC, which are calculated from the isothermal structural relaxation times and fitted to Eq. (14). Solid lines denote the fitting curves. (c) Scaling of the activation volume of BMPC in terms of Eq. (14) by using the values of its fitting parameters collected in panel (b). The linear fit represented by the solid line indicates the quality of the scaling.

temperature parametrization formulated by Eq. (21) is indeed reasonable for a reference state $(T_0, p_0)$, which can usually be chosen at ambient pressure. Therefore, as an example, we determine the activation volume for BMPC at the reference pressure $p_0=0.1$MPa. As can be seen in Fig. 2(c), the temperature dependence $\upsilon_{act}(T, p_0)$ can be successfully fitted to Eq. (20), although the activation volume unlike the specific volume decreases with increasing temperature in isobaric conditions, which can be easily observed for instance from the inspection of Fig. 1(b). As a consequence, values of the parameter $F_1$ in Eq. (20) are expected to be negative numbers.

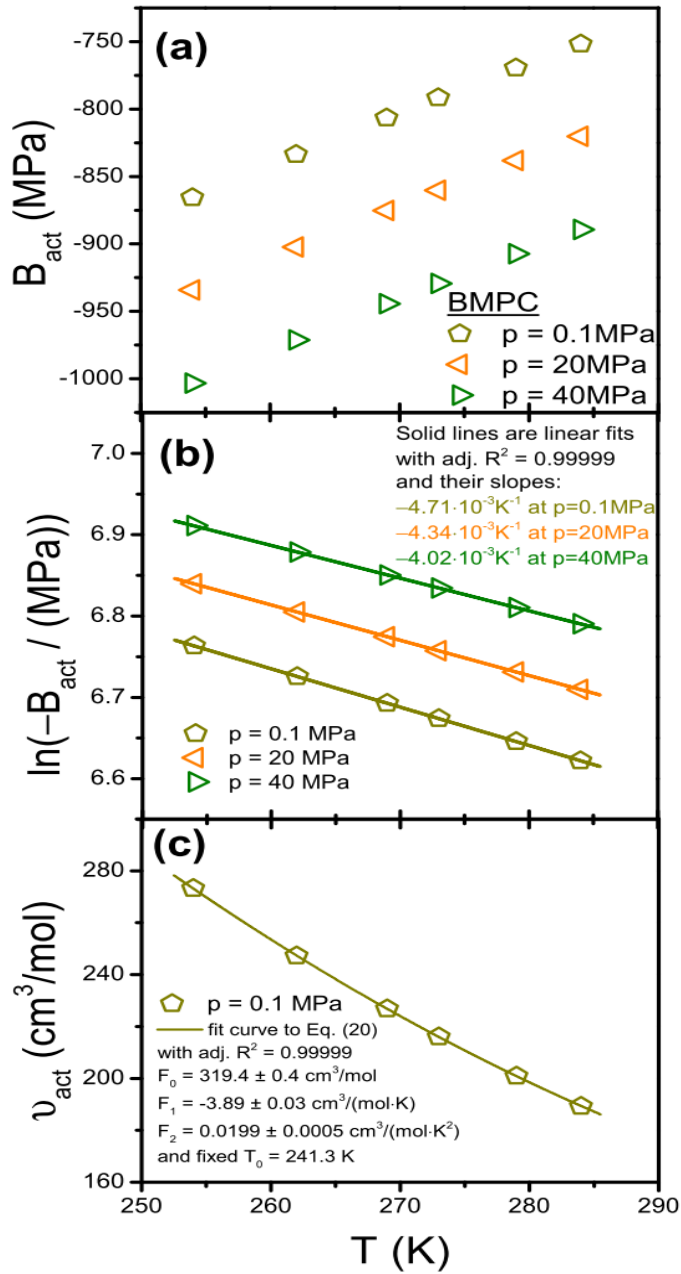

Fig. 2 (a) Plot of temperature dependences of the isothermal bulk modulus for the activation volume of BMPC in chosen isobaric conditions. (b) Plot of the dependences earlier shown in panel (a) with using a logarithmic representation of the isothermal bulk modulus for the activation volume and their linear fits denoted by solid lines. (c) Plot of the temperature dependence of the activation volume of BMPC at ambient pressure.

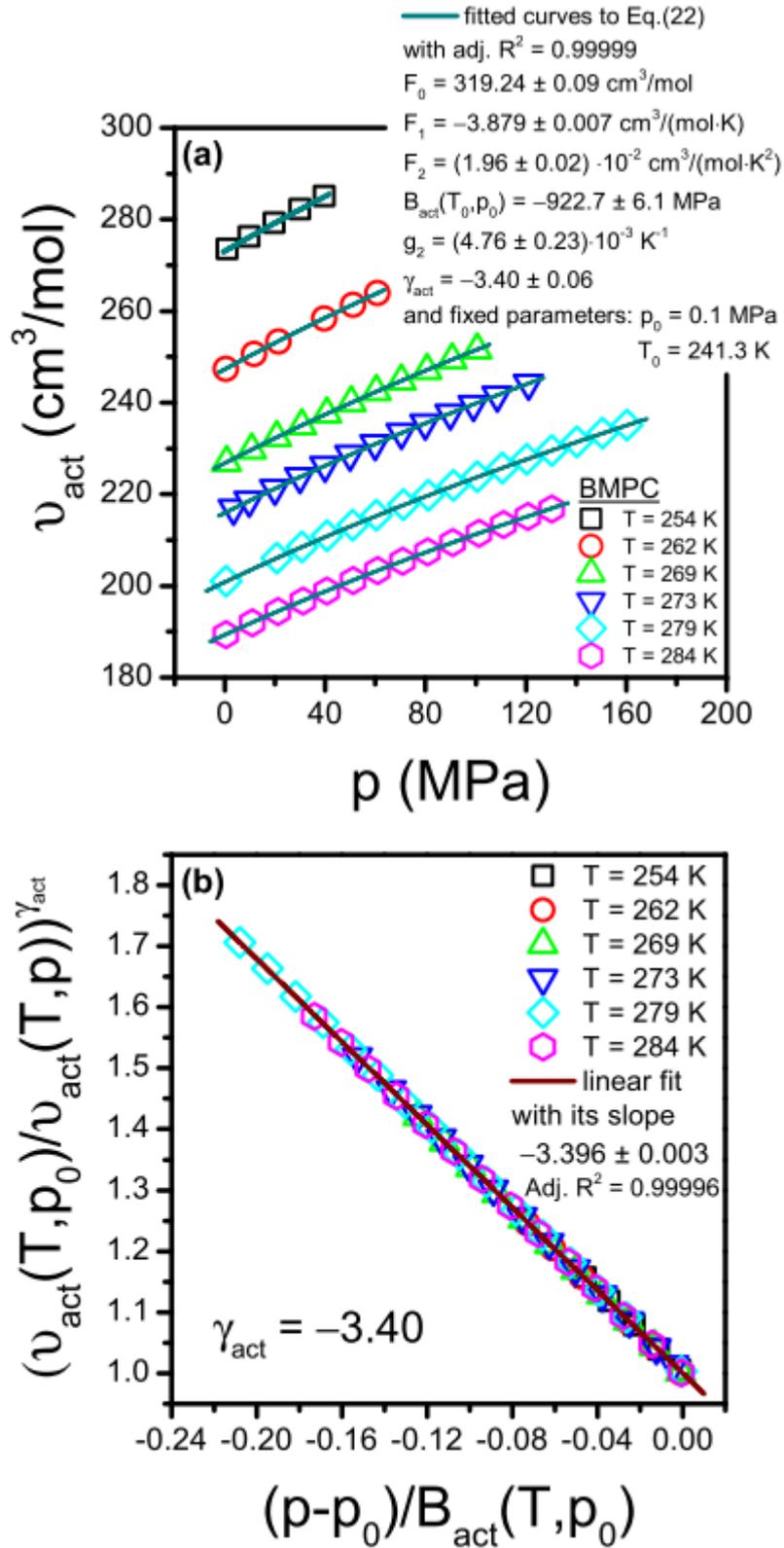

Fig. 3 (a) Plot of the pressure dependences earlier shown in Fig. 1(b) currently fitted to Eq. (22). Solid lines denote the fitting curves. (b) Scaling of the activation volume of BMPC in terms of Eq. (14) parametrized by Eqs. (20) and (21) with using the values of the fitting parameters collected in panel (a) for Eq. (22). The linear fit represented by the solid line indicates the quality of the scaling.

The preliminary tests of the temperature parametrizations of Eq. (14) expressed by Eqs. (20) and (21) show that Eq. (22) should be the appropriately generalized EOS for the activation volume. Thus, we exploit Eq. (22) in further analyses to check whether it is able to properly describe the activation volume considered as a function of pressure and temperature for glass formers from different material classes. First, we check how Eq. (22) approximates the pressure dependences of the activation volume of BMPC, which are shown in Fig. 1(b) and previously have been successfully fitted to the isothermal EOS given by Eq. (14). As can be seen in Fig. 3(a), the activation volume of BMPC can be also successfully fitted to the generalized EOS given by Eq. (22). Moreover, similarly to the performed scaling of the activation volume according to Eq. (14), we can use the values of the parameters found from the fitting procedure to Eq. (22) to construct a linear scaling plot for the activation volume (Fig. 3(b)), which is analogous to the PVT scaling plots earlier reported by us for the specific volume [39] analyzed in terms of the generalized version of Eq. (2) with appropriate temperature parametrizations $\upsilon(T, p_0)$ and $B_T(p_0)$. We achieve a very high quality scaling of the activation volume of BMPC with the value of the scaling exponent $\gamma_{act} \approx -3.40$, which is determined by fitting the activation volume data to Eq. (22) and is very close to that found from the isothermal EOS (Eq. (14)) for BMPC, i.e., $\gamma_{act} \approx -3.45$. In this way, the generalized EOS for the activation volume (Eq. (22)) and the scaling of the activation volume according to the EOS have been validated by using experimental data for the representative of van der Waal liquids.

Next, we exploit experimental data for representatives of polymers (PVAc) and protic ionic liquids (VH) to verify Eq. (22). We determine the activation volume of PVAc and VH by using isothermal structural relaxation times established from high pressure broadband dielectric measurements and fitted to Eq. (9). The fitting procedure performed herein according to Eq. (9) includes beside the isothermal structural relaxation times shown respectively in Figs. 4(a) and 5(a) for PVAc and VH also the isobaric structural relaxation times at ambient pressure in case of PVAc [50] and the isobaric structural relaxation times measured at ambient and high pressures in case of VH [52]. Exploiting Eq. (2) with the values of the EOS parameters earlier reported for PVAc [39] and herein established for VH [53] by using the earlier reported PVT measurement data [52], we can find a temperature-pressure function $\tau(\rho(T, p), T)$ for the fits to Eq. (9). The relaxation times and the fitted curves are shown as pressure functions for PVAc and VH in Figs. 4(a) and 5(a), respectively. Then, we evaluate the pressure dependences of the activation volume for each isotherm from the

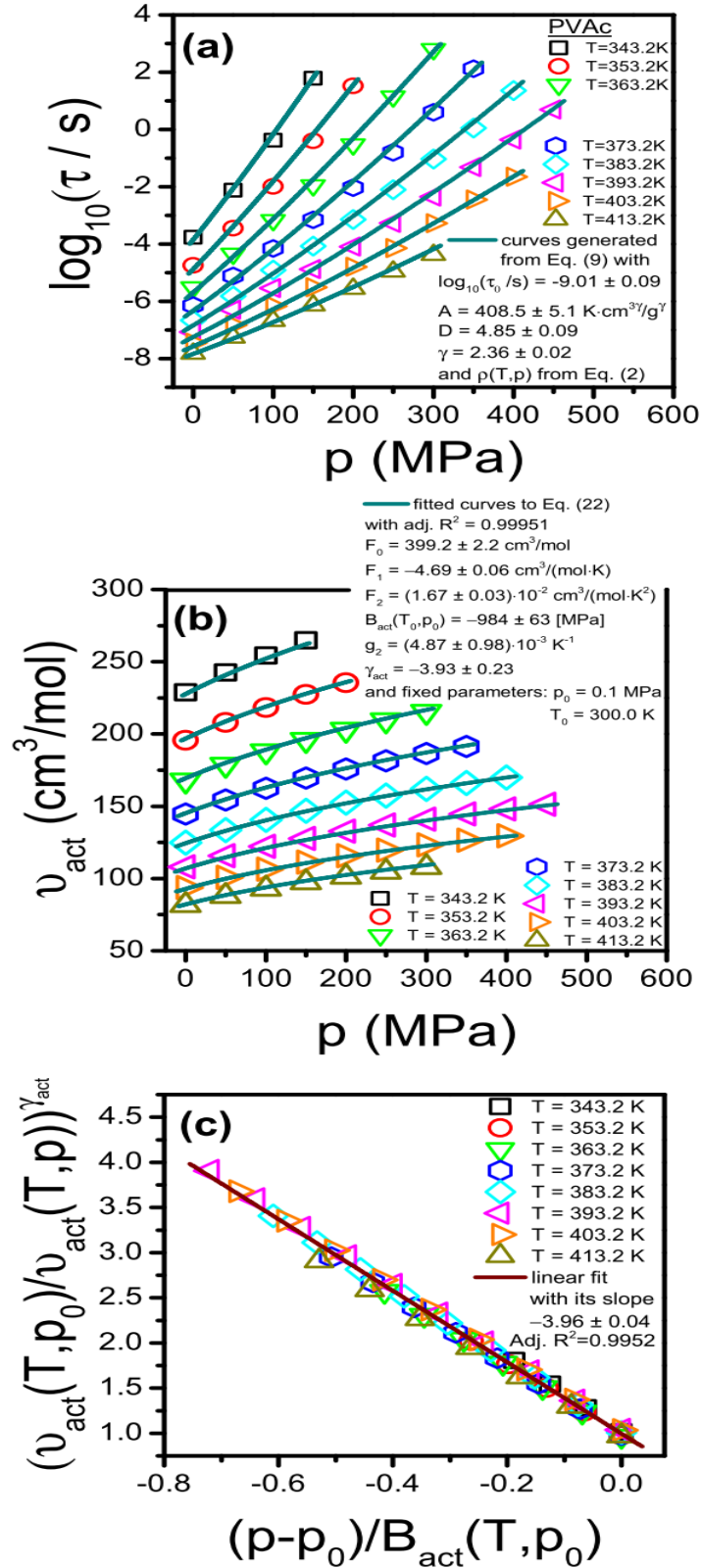

Fig. 4 (a) Plot of the pressure dependences of isothermal structural relaxation times τ of PVAc. Solid lines denote the curves generated from the shared fir of τ(T,ρ) to Eq. (9) and the temperature-pressure dependence of density established from Eq. (2) with the values of its parameters taken from [39] (b) Plot of the pressure dependences of the activation volume of PVAc, which are calculated from the isothermal structural relaxation times and fitted to Eq. (22). Solid lines denote the fitting curves. (c) Scaling of the activation volume of PVAc in terms of Eq. (14) parametrized by Eqs. (20) and (21) with using the values of the fitting parameters collected in panel (b) for Eq. (22). The linear fit represented by the solid line indicates the quality of the scaling.

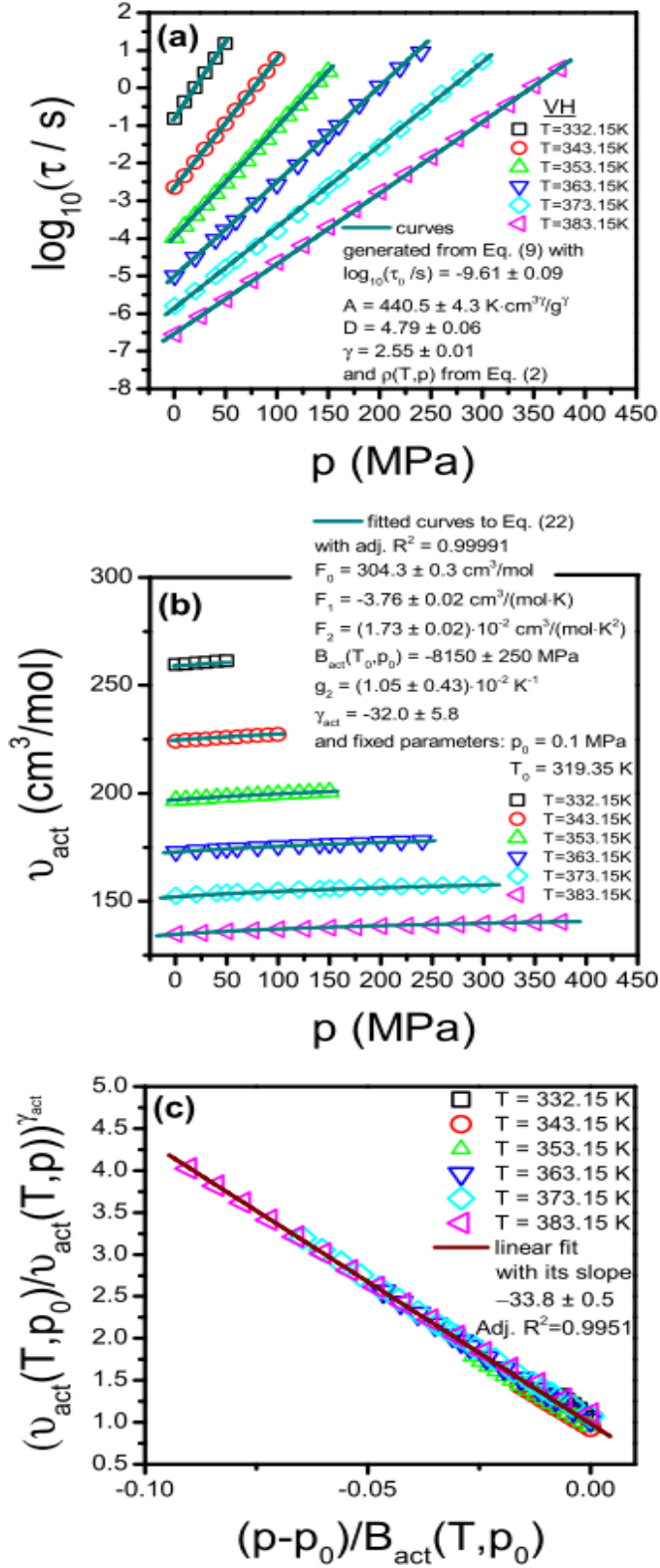

Fig. 5 (a) Plot of the pressure dependences of isothermal structural relaxation times τ of VH. Solid lines denote the curves generated from the shared fit of τ(T,ρ) to Eq. (9) and the temperature-pressure dependence of density established from Eq. (2) with the values of its parameters taken from [53] (b) Plot of the pressure dependences of the activation volume of VH, which are calculated from the isothermal structural relaxation times and fitted to Eq. (22). Solid lines denote the fitting curves. (c) Scaling of the activation volume of VH in terms of Eq. (14) parametrized by Eqs. (20) and (21) with using the values of the fitting parameters collected in panel (b) for Eq. (22). The linear fit represented by the solid line indicates the quality of the scaling.

definition given by Eq. (1). As can be seen in Fig. 4(b) for PVAc and Fig. 5(b) for VH, we can successfully fit the pressure dependences $\upsilon_{act}(p)$ to Eq. (22) at the fixed reference pressure $p_0$=0.1MPa. By using the values of the parameters found from the fitting procedure to Eq. (22), we can construct the linear scaling plots for the activation volume (Figs. 4(c) for PVAc and Figs. 5(c) for VH). The scaling quality is satisfactory in both the cases, although it is not perfect unlike that established for BMPC. The possible reasons for the slightly worse quality of the scaling of the activation volume seem to be different for PVAc and VH. In case of PVAc, one can expect that it is caused by using a rare set of PVT experimental data [51] to find the fitting values of the EOS parameters given by Eq. (2), which have been also suggested [39] as a reason for a slightly worse scaling of the specific volume in terms of Eq. (2). However, in case of VH, the quality of the scaling of $\upsilon_{act}$ seems to be a consequence of our previous finding [52] that this protic ionic liquid does not obey the thermodynamic scaling law perfectly. In this context, it is interesting to check whether it is possible to perform the discussed scaling of the activation volume for glass formers that do not obey the thermodynamic scaling law at all.

To answer the above question we consider experimental data for DPG, which is a strongly H-bonded liquid. We have previously confirmed [57] the isothermal and isobaric structural relaxation times of DPG cannot be superimposed on any curve that represents a function of $\rho^{\gamma}/T$, where $\gamma$ is a material constant. As an example, we have fitted all the structural relaxation times collected in different thermodynamic conditions to Eq. (9). It means that we have been trying to repeat the same procedure as that applied to structural relaxation times of BMPC, PVAc, and VH. As a result, we have obtained a very low quality fit. The curves generated from the values of this fit parameters are depicted by dashed lines in Fig. 6(a) as pressure functions found by using Eq. (2) with the values of the EOS parameters established by using the isobaric experimental data combined from ambient and high pressure PVT measurements [54,56,59]. Therefore, we have fitted separately each isotherm of structural relaxation times of DPG to Eq. (9) to be able to accurately evaluate the activation volume of DPG. Then, we have obtained high quality fits (see solid lines in Fig. 6(a)) and reliable values of the activation value from the definition given by Eq. (1). These values of $\upsilon_{act}$ for DPG have been successfully fitted to Eq. (22) as can be seen in Fig. 6(b), and later quite satisfactorily scaled (Fig. 6(c)) with the same method as that applied to scale the activation volume of earlier

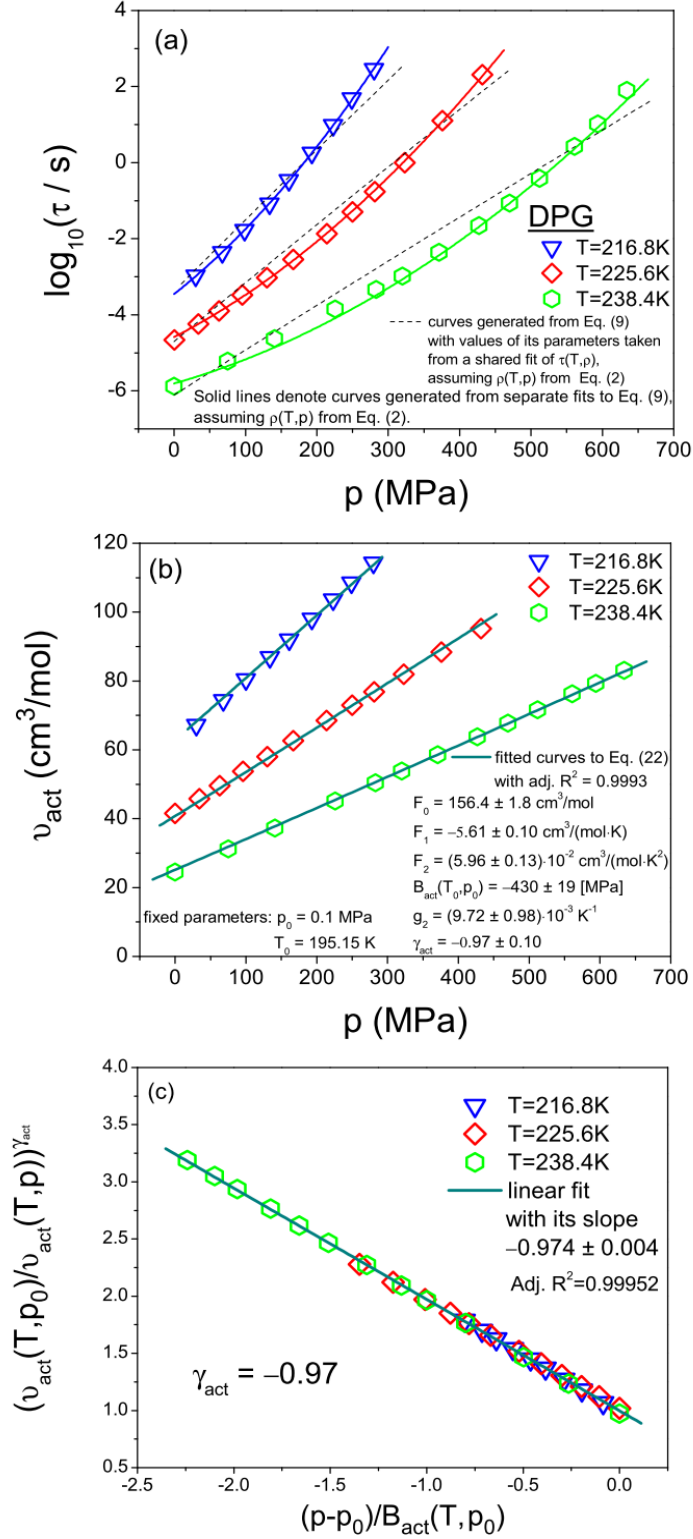

Fig. 6 (a) Plot of the pressure dependences of isothermal structural relaxation times τ of DPG. Dashed and solid lines denote the curves generated respectively from the shared fit and the separate isothermal fits of τ(T,ρ) to Eq. (9) and the temperature-pressure dependence of density established from Eq. (2) with the values of its parameters taken from [59] (b) Plot of the pressure dependences of the activation volume of DPG, which are calculated from the separately fitted isothermal structural relaxation times and fitted to Eq. (22). Solid lines denote the fitting curves. (c) Scaling of the activation volume of DPG in terms of Eq. (14) parametrized by Eqs. (20) and (21) with using the values of the fitting parameters collected in panel (b) for Eq. (22). The linear fit represented by the solid line indicates the quality of the scaling.

considered materials. This result shows that the activation volume can be scaled in terms of Eq. (14) parametrized by Eqs. (20) and (21), even if the structural relaxation times of a given material do not obey the thermodynamic scaling law. This finding can be explained by the already mentioned fact that Eq. (14) as well as their supplementations with Eqs. (20) and (21) are derived without any assumption that the power-law density scaling is valid.

Since Eq. (22) has turned out to be the proper EOS for the activation volume of glass formers that belong to various material groups, including also materials that do not obey the thermodynamic scaling law in the exact sense, we consider our MD simulation data [40] obtained for the model KABLJ supercooled liquid to verify this EOS. It has been confirmed [27,40] that the scaling exponent $\gamma$ depends on density for the KABLJ system. However, it is possible to determine an effective value of the scaling exponent $\gamma$. It has been achieved by us [40] by fitting structural relaxation times to Eq. (9), which results in $\gamma \approx 4.87$ for structural relaxation times expressed in the Lennard-Jones (LJ) potential units. Nevertheless, to accurately evaluate the activation volume for the KABLJ liquid, we fit separately each isotherm of $\tau^*$ (the star denotes $\tau$ in the LJ units) to Eq. (9). The isothermal structural relaxation times $\tau^*$ and their fits to Eq. (9) are shown in Fig. 7(a) as pressure functions by exploiting values of pressure calculated from our MD simulation data for the KABLJ system. Then, we can determine proper values of the activation volume for the KABLJ system and fit them to Eq. (22). As can be seen in Fig. 7(b), the pressure dependences of $\upsilon_{act}$ for the KABLJ supercooled liquid are very well described by the fitting curves to Eq.. (22). One can observe an interesting difference in the values of the parameter $g_2$ between considered real systems and the KABLJ model. For the latter, the value $g_2 < 0$, whereas the found values $g_2$ are positive for the real materials. According to Eq. (21), it means that the isothermal bulk modulus for the activation volume, which is negative, decreases with increasing temperature in isobaric conditions in the case of KABLJ liquid, which is just the opposite of that for the real glass formers. Nevertheless, the scaling of the activation volume obtained for the model system is satisfactory also with a negative value of the scaling exponent $\gamma_{act}$ (Fig. 7(c)) similarly to those found for real materials. Additionally, it should be noted that the scaling procedure of the activation volume for both real and model systems does not require using the reduced units suggested by Dyre's group within the theory of isomorphs [60], because Eq. (14) has a quotient character similarly to Eq. (2) for which the issue has been discussed by us in [40].

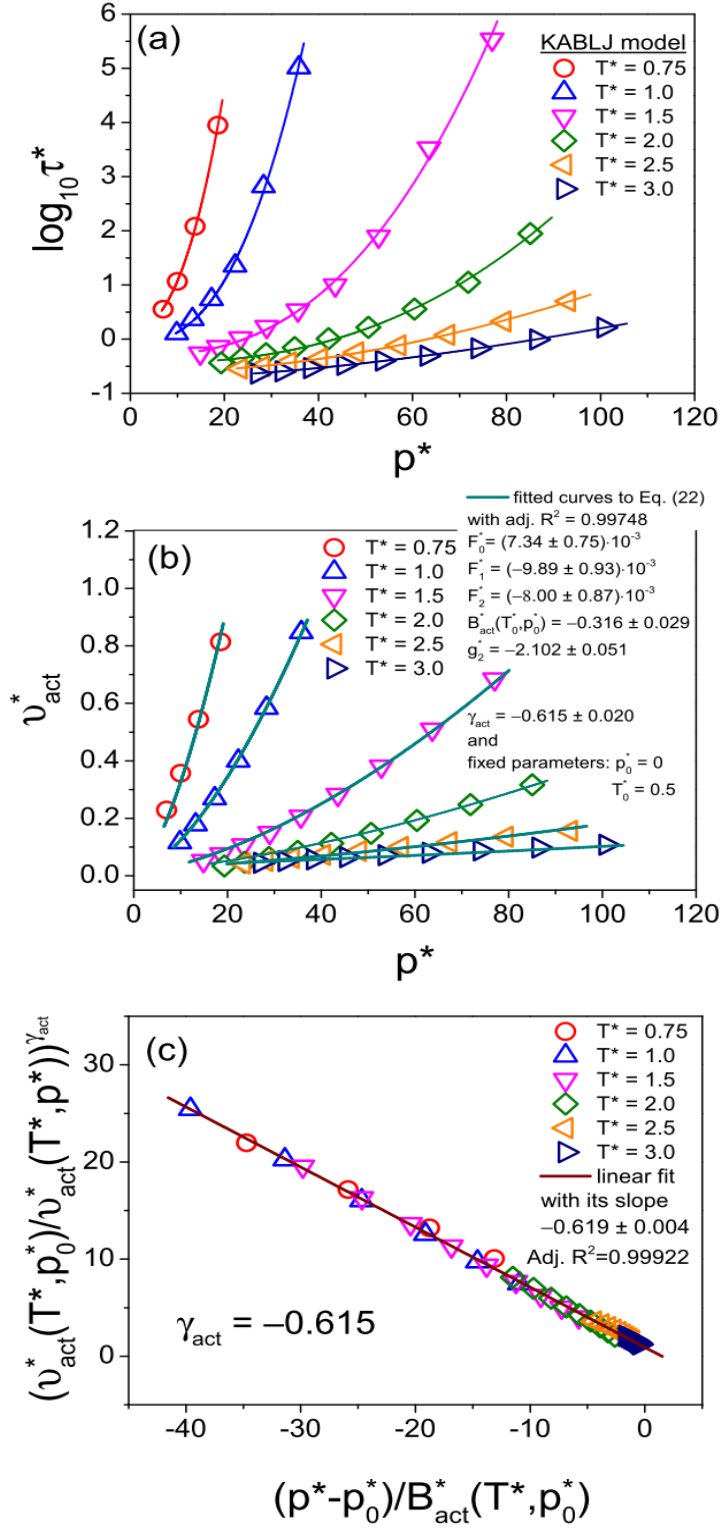

Fig. 7 The star symbol in each panel indicate that a given quantity is expressed in LJ units. (a) Plot of the pressure dependences of isothermal structural relaxation times $\tau^*$ of the KABLJ liquid. Solid lines denote the curves generated from the separate isothermal fits of $\tau^*(T^*,\rho^*)$ to Eq. (9) and the temperature-pressure dependence of density established from Eq. (2) with the values of its parameters taken from [40] (b) Plot of the pressure dependences of the activation volume of the KABLJ liquid, which are calculated from the separately fitted isothermal structural relaxation times and fitted to Eq. (22). Solid lines denote the fitting curves. (c) Scaling of the activation volume of the KABLJ liquid in terms of Eq. (14) parametrized by Eqs. (20) and (21) with using the values of the fitting parameters collected in panel (b) for Eq. (22). The linear fit represented by the solid line indicates the quality of the scaling.

At the end of the discussion on the verification of Eq. (22) by using experimental and simulation data, we analyze the values of the scaling exponent $\gamma_{act}$ obtained from fitting the activation volumes to Eq. (22) in comparison with those calculated from Eq. (17) that should predict the values of $\gamma_{act}$ according to our theoretical discussion presented in Section III. According to this equation, we can calculate the value of $\gamma_{act}$ for a given material if we know the value of the parameter $\gamma_{EOS}$ found from fitting PVT data to Eq. (2) and the values of the parameters $\gamma$ and $D$ found from fitting e.g. structural relaxation times measured at different thermodynamic conditions to Eq. (9). We have applied this method to estimate values of the parameter $\gamma_{act}$ for each considered system. A comparison of the results established from this calculations based on Eq. (17) to those found by fitting the activation volumes to Eq. (22) has turned out to be highly satisfactory (see Table 1). Only in case of DPG, Eq. (17) has failed, but it is quite reasonable, because DPG is strongly hydrogen bonded liquid that does not obey the thermodynamic scaling law as it has been already mentioned and presented in Fig. 6(a) from which it can be easily observed that the shared fit to Eq. (9) represented by dashed lines weakly describes structural relaxation times of DPG. Thus, the effective values of $\gamma$ and $D$ found by this fit are not reliable and the value of $\gamma_{act}$ calculated from Eq. (17) by using these values of the parameters $\gamma$ and $D$ is also improper. For other systems, the values of $\gamma_{act}$ estimated from Eq. (17) are in accord with those found by fitting the activation volumes to Eq. (22). As can be seen in Table 1, the best prediction of the values $\gamma_{act}$ has been achieved for van der Waals liquids (PDE,BMPC) and the polymer PVAc. For the protic ionic liquid VH, for which the thermodynamic scaling is not perfect [52], the estimation of $\gamma_{act}$ from Eq. (17) is slightly worse and its standard deviation of determination is quite big, but it is still a good approximation of $\gamma_{ac}$ for this material. It is interesting that Eq. (17) can be successfully used to predict the value of $\gamma_{ac}$ in case of the KABLJ model, which does not obey the power law density scaling in the exact sense. It is a consequence of the fact that the effective values of the parameters $\gamma$ and $D$ established for the model system are reasonable, because it is possible to perform the thermodynamic scaling of the structural relaxation times of the KABLJ liquid to a good approximation with using this value of the scaling exponent $\gamma$ [40].

Table. 1 Values of the scaling exponent $\gamma_{act}$ found from fitting the activation volume to Eq. (22) compared with those calculated from Eq. (17) by using the parameters $\gamma$, $D$, and $\gamma_{EOS}$.

| System | $\gamma$ fitted by using Eq. (9) [a] | $D$ fitted by using Eq. (9) [a] | $\gamma_{EOS}$ fitted by using Eq. (2) | $\gamma_{EOS} - \gamma D$ | $\gamma_{act}$ calculated from Eq. (17) | $\gamma_{act}$ fitted by using Eq. (22) |
|---|---|---|---|---|---|---|
| PDE | 4.42 ± 0.02 | 4.25 ± 0.04 | 9.51 ± 0.04 [b] | −9.28 ± 0.20 | −1.03 ± 0.03 | −1.02 ± 0.03 |
| BMPC | 8.00 ± 0.09 | 2.05 ± 0.03 | 12.69 ± 0.10 [c] | −3.68 ± 0.31 | −3.45 ± 0.29 | −3.40 ± 0.06 |
| PVAc | 2.36 ± 0.02 | 4.85 ± 0.09 | 9.19 ± 0.21 [b] | −2.26 ± 0.32 | −4.06 ± 0.63 | −3.93 ± 0.23 |
| VH | 2.55 ± 0.01 | 4.79 ± 0.06 | 11.79 ± 0.07 [d] | −0.42 ± 0.18 | −28 ± 12 | −32.0 ± 5.8 |
| DPG | 1.99 ± 0.02 | 4.94 ± 0.15 | 10.23 ± 0.03 [e] | 0.43 ± 0.32 | 23 ± 17 | −0.97 ± 0.10 |
| KABLJ model | 4.84 ± 0.13 [f] | 2.42 ± 0.18 [f] | 4.58 ± 0.19 [f] | −7.13 ± 0.95 | −0.64 ± 0.10 | −0.62 ± 0.02 |

[a] For the considered real glass formers, the values of $\gamma$ and $D$ have been established herein by fitting $\tau(\rho,T)$ to Eq. (9) in the entire T-p dielectric experimental range, exploiting $\rho(T,p)$ from Eq. (2).
[b] taken from Ref. 39
[c] taken from Ref. 58
[d] taken from Ref. 53
[e] taken from Ref. 59
[f] taken from Ref. 40

Since Eq. (17) is validated in case of all considered systems except DPG for which the deviation from the thermodynamic scaling law is most pronounced, it is worth noting that Eq. (17) combined with Eq. (18) provides an useful method for predicting the pressure dependence of the isothermal bulk modulus for the activation volume, which is based on Eq. (15). The latter equation with Eqs. (17) and (18) enables us to predict the dependences $B_{act}(p)$ with calculating no activation volume.

*(c) Bulk Modulus for the Activation Volume in the Density Scaling Regime*

As a natural application of the EOS formulated by us for $\upsilon_{act}$ in Section III, we discuss the isothermal bulk modulus for the activation volume in terms of Eq. (22). We have argued in Section III that Eq. (14) implies a linear pressure dependence $B_{act}(p)$, the slope and intercept of which can be predicted by Eqs. (17) and (18), respectively. Taking into account the isothermal precursor (Eq. (14)) of Eq. (22), we can expect that all the findings established for Eq. (14) should be valid in case of Eq. (22). To verify this hypothesis we exploit the data for all the earlier considered real and model systems, but the detailed analysis of the issue is first performed by using experimental data for PDE, which is a prototypic van der Waals liquid. According to the procedure valid for materials that obey the thermodynamic scaling law, the isothermal and isobaric structural relaxation times of PDE [48] are fitted to Eq. (9) to find one set of its parameters values. The pressure dependences of the isothermal structural relaxation times for PDE and their fits generated from Eq. (9) are shown in Fig. 8(a) as pressure functions determined by using Eq. (2) with the earlier reported [39] values of the EOS parameters. Then, we calculate the values of the activation volume of PDE from the definition given by Eq. (1) and fit them to Eq. (22) as shown in Fig. 2(b). After numerical differentiating the fits with respect to pressure we can find the pressure dependences of the isothermal bulk modulus for the activation volume according to its definition given by Eq. (13). As can be seen in Fig, 8(d), the dependences $B_{act}(p)$ can be very well described by pressure linear functions with the same value of their slope that equals $-1.008\pm0.002$, which is very close the scaling exponent $\gamma_{act} = -1.02\pm0.03$ found from fitting the activation volume of PDE to Eq. (22). It means that the linear function $B_{act}(p)$ given by Eq. (15), which is a consequence of the EOS (Eq. (14)) for the activation volume, is validated by experimental data of PDE.

For comparison, in Fig. 8(c), we show the pressure dependences of the isothermal bulk modulus $B_T(p)$ for the specific volume of PDE determined at the same temperatures at which the dependences $B_{act}(p)$ have been evaluated. The found dependences $B_T(p)$ are also linear and characterized by the same value of their slope $-9.514\pm0.004$, which is identical to that found [39] from fitting the activation volume of PDE to Eq. (2). This results is in accord with Eq. (3), which follows from the EOS (Eq. (2)) for the specific volume, and confirms that the PVT are taken from the sufficiently low compressibility region required to use Eq. (2).

We can also compare the scaling plots for the specific volume of PDE (Fig. 8(e)) and for the activation volume of this prototypical van der Waals liquid (Fig. 8(f)), which are performed respectively in terms of Eq. (2) with its temperature parametrizations and Eq. (14) supplemented with Eqs. (20) and (21) by using the values of the scaling exponents $\gamma_{EOS}$ and $\gamma_{act}$ found from fitting the specific volume to Eq. (2) and the activation volume to Eq. (22). Moreover, on the basis of Eqs. (2), (14), (16), and (17), one can expect that the scaling in terms of Eq. (2) is maintained if we replace $\upsilon$ with $\upsilon_{act}$ and $\gamma_{EOS}$ with $\gamma_{act}$, whereas the scaling in terms of Eq. (14) is held if we perform a reverse transformation, i.e., $\upsilon_{act} \to \upsilon$ and $\gamma_{act} \to \gamma_{EOS}$. As can be seen in Figs. 8(e) and 8(f) after taking into account their right ordinate axes, the scaling plots remain indeed unchanged if we make such transformations. Since the validity of the relationship between $\gamma_{act}$ and $\gamma_{EOS}$ given by Eq. (17) has been confirmed (see Table 1), the correspondence $(\upsilon_{act}(T,p_0)/\upsilon_{act}(T,p))^{\gamma_{act}} = (\upsilon(T,p_0)/\upsilon(T,p))^{\gamma_{EOS}}$ established for PDE in Figs. 8(e) and 8(f) also gives evidence of the validity of Eq. (11), which can be applied to predict the activation volume at high pressures by using values of the activation volume at a reference pressure and the specific volume at given T and p if values of the parameters $\gamma_{EOS}$, $\gamma$, and D are known. As a consequence, Eq. (12) is also validated for PDE.

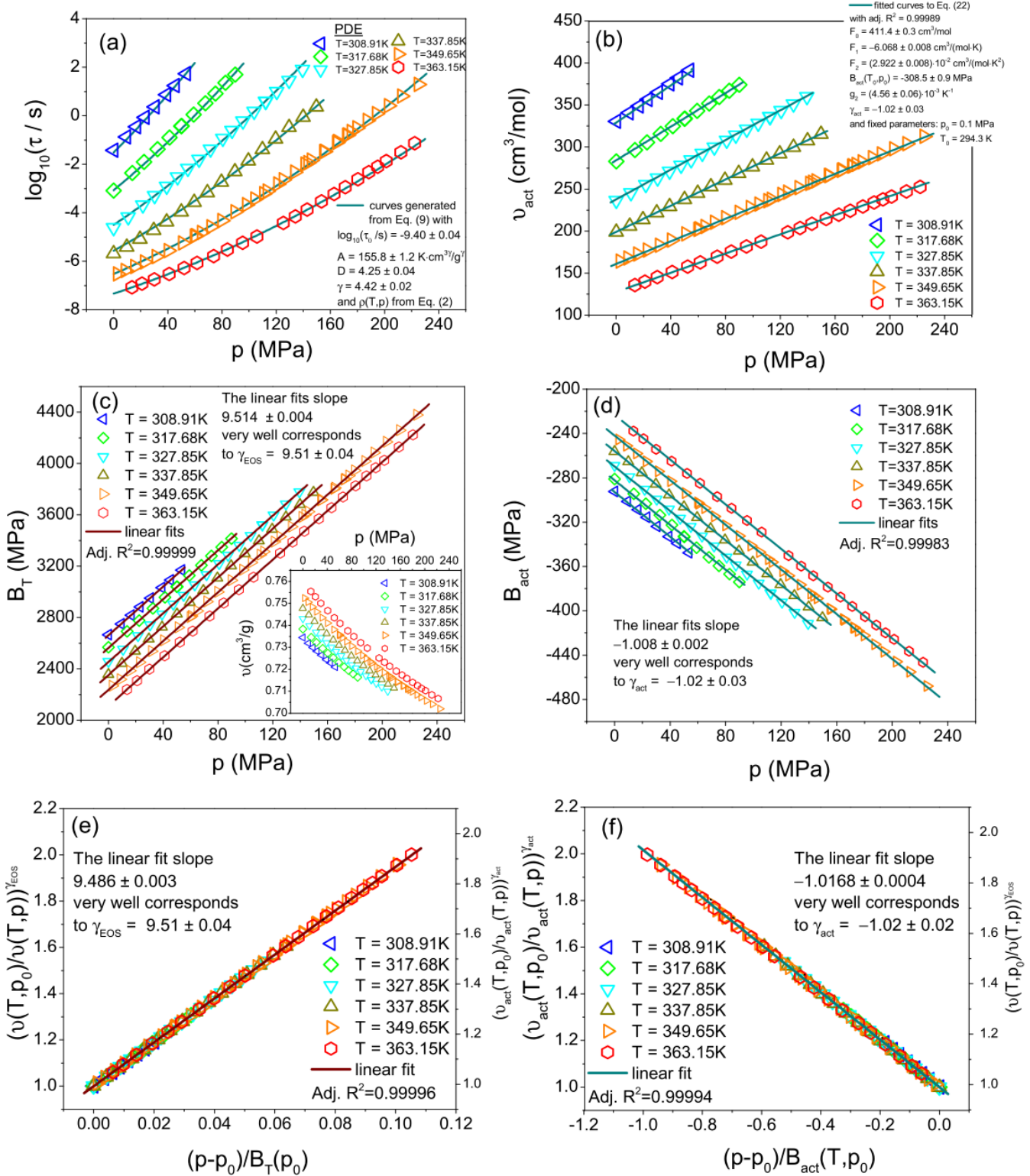

Fig. 8 (a) Plot of the pressure dependences of isothermal structural relaxation times τ of PDE. Solid lines denote the curves generated from the shared fit of $\tau(T,\rho)$ to Eq. (9) and the temperature-pressure dependence of density established from Eq. (2) with the values of its parameters taken from [39] (b) Plot of the pressure dependences of the activation volume of PDE, which are calculated from the isothermal structural relaxation times and fitted to Eq. (22). Solid lines denote the fitting curves. (c) Plot of the pressure dependences of the isothermal bulk modulus for the specific volume of PDE, which are determined from the isothermal pressure dependences of the specific volume shown in the inset. The linear fits of the dependences, which are denoted by solid lines, indicate that the slopes of the linear fits are the same and correspond very well to the scaling exponent $\gamma_{EOS}$ found [39] from fitting PVT data for PDE to Eq. (2). (d) Plot of the pressure dependences of the isothermal bulk modulus for the activation volume of PDE, which are determined from the isothermal pressure dependences of the activation volume shown in panel (b). The linear fits of the dependences, which are denoted by solid lines, indicate that the slopes of the linear fits are the same for a given materials and correspond very well to the scaling exponent $\gamma_{act}$ found from fitting of the activation volumes to Eq. (22). (e) Scaling of the specific volume of PDE in terms of Eq. (2) with using the values of its fitting parameters reported in [39]. The linear fit represented by the solid line indicates the quality of the scaling. (f) Scaling of the activation volume of PDE in terms of Eq. (14) parametrized by Eqs. (20) and (21) with using the values of the fitting parameters collected in panel (b) for Eq. (22). The linear fit represented by the solid line indicates the quality of the scaling. The right axes of panels (e) and (f) indicate the possible equivalent quantities that match up the scaling of the quantities presented on the left axes of the panels.

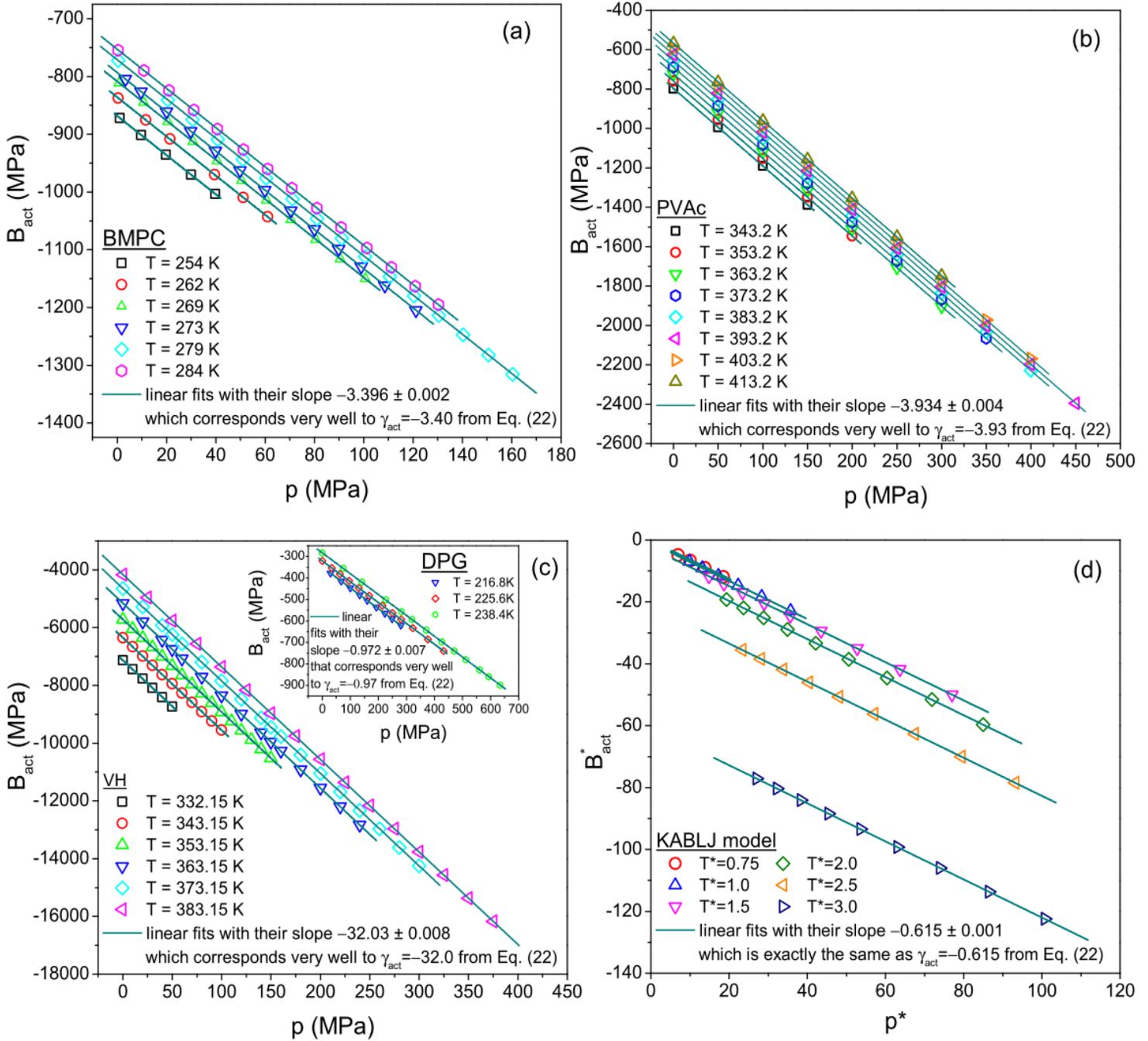

Fig. 9 Plots of the pressure dependences of the isothermal bulk modulus for the activation volume for (a) BMPC, (b) PVAc, (c) VH, (the inset in panel (c)) DPG, and (d) the KABLJ model. The linear fits of the dependences, which are denoted by solid lines, indicate that the slopes of the linear fits are the same for a given materials and correspond very well to the scaling exponent $\gamma_{act}$ found from fitting of the activation volumes to Eq. (22).

It should be noted that the obtained increasing pressure functions $B_T(p)$ are an expected result in contrast to the decreasing pressure dependences $B_{act}(p)$. The latter case is extremely interesting, because the previous analyses of the activation volume of glass formers suggested the curvature of plots of the pressure dependences of the activation volume [9], which could rather result in the increasing pressure dependences of $B_{act}$. Therefore, we have tested the dependences $B_{act}(p)$ for other considered systems by following the same procedure as that used in the case of PDE. It means we have calculated the isothermal bulk modulus for the activation volume from its definition given by Eq. (13), exploiting numerical differentiations of the fits of $\upsilon_{act}(p)$ to Eq. (22) for each investigated system. As a result, we have established the pressure dependences of $B_{act}$, the plots of which are demonstrated in Fig. 9 for BMPC, PVAc, VH, DPG, and the KABLJ model. For the latter (Fig. 9(d)), one can observe the already mentioned behavior of the isothermal bulk modulus for the activation volume, which indeed decreases with increasing temperature in isobaric conditions contrary to that for real materials (Figs. 9(a)-(c), and 8(d)). However, in case of each system, the found dependences $B_{act}(p)$ are linear to a very good approximation and can be characterized by the slopes of their plots, which are constant for a given material and correspond very well to the value of the scaling exponent $\gamma_{act}$ determined by fitting the activation volumes of the material to the EOS given by Eq. (22). The established agreement between the values of $\gamma_{act}$ and the slope of the linear plot of the dependence $B_{act}(p)$ shows that the surprising decrease in $B_{act}$ with increasing pressure is a consequence of the relation suggested by Eq. (17). From inspection of the earlier reported [13,36-39,43,45,61] and current (see Table 1) results of the analyses based on Eqs. (2) and (9) one can claim that the values of the parameters that constitute the right side of Eq. (17) meet the condition $\gamma_{EOS} < \gamma D$ in case of all previously and herein tested materials that obey the thermodynamic scaling law at least to a sufficiently good approximation. Therefore, the theoretical considerations given to $\upsilon_{act}$ and $B_{act}$ in Section III lead to the conclusion that the linear dependence $B_{act}(p)$ defined by Eq. (15) is expected to decrease with increasing pressure due to $\gamma_{act} < 0$ for many glass formers, including also model systems like the KABLJ liquid. It is worth noting that although the condition $\gamma_{EOS} < \gamma D$ is not met in case of DPG (see Table 1), the scaling exponent $\gamma_{act}$ determined from experimental data for this strongly hydrogen bonded liquid is also negative. It means that the decreasing pressure dependence of the isothermal bulk modulus for the activation volume may be a common rule for glass forming materials, however, Eq. (17) is able to predict reliably the values of $\gamma_{act}$ only in the thermodynamic scaling regime. Eq. (17) should be also a good estimator for $\gamma_{act}$ if it is possible to find proper effective values of the scaling exponent $\gamma$ and the parameter D from Eq. (9) as it is in case of VH and the KABLJ model.

Finally, another isothermal EOS for the activation volume should be discussed. Papathanassiou and Sakellis have recently formulated this EOS (Eq. (9) in [62]) on the assumption that $B_{act}(T,p)/B_{act}(T,p_0)$ at a given $T$ is equal to $B_T(p)/B_T(p_0)$, which relies on the earlier simplifying suggestion, $|B_{act}| \approx B$, based on a more general assumption that $|B_{act}| \approx B/\lambda$, where $0 < \lambda < 1$ [63]. Taking into account the theory proposed in this paper for the activation volume defined by Eq. (1) (which is different from that exploited in [62]) and the isothermal bulk modulus for the activation volume defined by Eq. (13) as well as the obtained results of experimental and simulation tests of the proposed herein isothermal and generalized equations of state for $\upsilon_{act}$ and the related description of $B_{act}(p)$, we need to note that the approach reported in [62,63] is not appropriate to properly describe the pressure dependences $\upsilon_{act}$ and $B_{act}$ determined herein in case of most considered systems. First, the parameter $\lambda$ that can be expressed on the basis of Eq. (18) as follows $\lambda = |\gamma_{EOS} - \gamma D|$. This implies that the assumption given by the condition $0 < \lambda < 1$ is not sufficient, because it is met (see Table 1) only for VH (and DPG, but the strongly hydrogen bonded liquids should be excluded from this analyses, because we have already shown that Eq. (17) and consequently Eq. (18) are not valid for DPG). The other examined systems are characterized by the values of $\lambda$ considerably greater than 1. Nevertheless, Eq. (9) in [62] has a quotient character and the mentioned problem does not affect critically this EOS for the activation volume, and consequently it could be neglected if there were no other discrepancies between this EOS predictions and the activation volumes and their scaling, which have been established herein for the real and model systems. However, this crucial discrepancy can be observed for all systems tested by us. To discuss the problem, it should be pointed out that Eq. (14) implies $\gamma_{act} = (\partial B_{act}(T,p)/\partial p)_T$ at $p=p_0$, which can be easily seen from Eq. (15). Then, by comparison Eq. (9) in [62] with Eq. (14), one can find that the parameter $\gamma_{act}$ simply corresponds to the derivative $(\partial B_T(p)/\partial p)_T$ at $p=p_0$ in [62]. To meet the discussed assumption made in [62], the straightforward relation, $\gamma_{act} = \gamma_{EOS}$, should be valid according to Eqs. (2) and (3). It means that the slopes of the linear pressure dependences of $B_{act}$ and $B_T$ should be approximately the same, but it is not the case as can be seen in Figs. 8(e)-(f) and 9 and especially in Table 1 in which the values of $\gamma_{act}$ and $\gamma_{EOS}$ have been collected.

This finding has been already discussed by us in terms of Eq. (17), which shows that the relation, $\gamma_{act} = \gamma_{EOS}$, is not valid in the thermodynamic scaling regime also considered in [62].

**V. Summary and Conclusions**

In this paper, the activation volume and the isothermal bulk modulus have been considered in the thermodynamic scaling regime. Although the main course of the presented discussion relies on the assumption that the power law density scaling of structural relaxation times or viscosities is valid, we have derived the isothermal equation of state for the activation volume, which does not require obeying the thermodynamic scaling law, but only demands the region of small compressibility for the activation volume. As a consequence, the formulated isothermal EOS (Eq. (14)) implies the linear pressure dependence of the isothermal bulk modulus for the activation volume (Eq. (15)), which is in accord with the pressure dependences $B_{act}(p)$ established by using experimental and simulation data. As a key result, we have formulated the generalized EOS for the activation volume (Eq. (22)) by exploiting suggested temperature parametrizations of the isothermal EOS (Eqs. (20) and (21)), which have been confirmed by using experimental data.

After the successful test of using the isothermal EOS in fitting the pressure dependences of the activation volume of BMPC, which is a typical van der Waals liquid, the generalized equation of state for the activation volume have been very successfully applied to describe the pressure dependences of the activation volume determined by using structural relaxation times of glass formers that belong to various material groups such as van der Waals liquids, polymers, protic ionic liquids, and strongly hydrogen bonded liquids. Moreover, we have checked that the activation volumes found from structural relaxation times estimated from simulation data for the KABLJ model can be very well fitted to the EOS for the activation volume.

The following important conclusions can be drawn by using the equations of state for the activation volume. (i) The isothermal EOS for the activation volume predicts some kind of scaling for the activation volume. We have confirmed that the scaling of the activation volume can be performed for each considered real and model system by using the value of the scaling exponent $\gamma_{act}$, which can be found from fitting the activation volumes to the generalized EOS or its isothermal precursor, independently of that whether a given system obeys the thermodynamic scaling law or not. The same property has been earlier established [38,39] for the scaling of the specific volume with the value of the scaling exponent $\gamma_{EOS}$ found from fitting PVT data to the EOS for the specific volume (Eq. (2)). (ii) There is a relation given by Eq. (17) between the scaling exponent $\gamma_{act}$ and the exponents $\gamma_{EOS}$ and $\gamma$, where the latter is the exponent that enables us to scale structural relaxation times or viscosities according to the power law density scaling law. The relation is valid in case of systems, the molecular dynamics of which obeys the scaling law. (iii) As already mentioned the EOS for the activation volume leads to the linear pressure dependences of the isothermal bulk modulus for the activation volume (Eq. (15)). By analyzing the experimental and simulation data, we have found that the isothermal bulk modulus for the activation volume linearly decreases with increasing pressure. This surprising result can be rationalized by Eq. (17), which shows that the sign of the scaling exponent $\gamma_{act}$ is the same as that of the following expression, $\gamma_{EOS} - \gamma D$, where $D$ is one of the parameters of Eq. (9). The negative values of $\gamma_{act}$ can be explained by the finding that $\gamma_{EOS} < \gamma D$ in case of all considered systems except for the strongly hydrogen bonded liquid DPG, for which the thermodynamic scaling is not valid. (iv) By comparing the EOS for the activation volume (Eq. (14)) with the EOS for the specific volume (Eq (2)), we have shown that the inverse reduced activation volume can be expressed by the scaled inverse reduced specific volume or the scaled inverse reduced isothermal bulk modulus for the specific volume (Eqs. (11) and (12)). These findings enable us to distinguish the strong volumetric contribution to the activation volume from the dynamic hallmark that is reflected in the scaling exponents of Eqs. (11) and (12), in which the parameter $\gamma$ plays an important role as the exponent of the power law density scaling for molecular dynamics near the glass transition. (v) As a consequence, these equations of state give us an interesting convenient opportunity to predict the pressure dependences of $B_{act}$ from Eq. (15) with Eqs. (17) and (18) by using structural relaxation times or viscosities combined with PVT data, omitting the activation volume evaluation. Similarly, if we know the activation volumes in the ambient pressure limit, the values of the parameters of the EOS for the specific volume (Eq. (2)) as well as the values of the parameters $\gamma$ and D found from fitting structural relaxation times or viscosities data to Eq. (9), we can reproduce the activation volumes for various temperature-pressure conditions by exploiting Eq. (11), which is a reliable procedure in case of materials that obey the thermodynamic scaling law at least to a good approximation.

The presented applications and predictions based on the proposed formalism for the activation volume and the isothermal bulk modulus for the activation volume indicate that it is expected to be a very useful tool to investigate interrelations between important parameters that characterize properties of glass forming materials such as the activation volume, the fragility parameter, the pressure coefficient of the glass transition, and the length scale of the spatially heterogeneous dynamics near the glass transition, taking into consideration the whole thermodynamic space.

**Acknowledgements**

A.G., K.K., K.G., and M.P. gratefully acknowledge the financial support from the Polish National Science Centre within the program MAESTRO 2. K.K is deeply thankful for the stipend received within the project "DoktoRIS - the stipend program for the innovative Silesia", which is co-financed by the EU European Social Fund.

**References**

[1] G. Floudas, M. Paluch, A. Grzybowski, K. Ngai, *Molecular Dynamics of Glass-Forming Systems: Effects of Pressure*, (Series: *Advances in Dielectrics*, Series Editor: Friedrich Kremer), Springer-Verlag Berlin Heidelberg 2011

[2] C. M. Roland, S. Hensel-Bielowka, M. Paluch, and R. Casalini, Rep. Prog. Phys. **68**, 1405 (2005)

[3] M. Paluch, A. Patkowski, and E. W. Fischer, Phys. Rev. Lett. **85**, 2140 (2000)

[4] S. Hensel-Bielowka and M. Paluch, Phys. Rev. Lett. **89**, 025704 (2002)

[5] M. Paluch, M. Sekula, S. Pawlus, S. J. Rzoska, J. Ziolo, and C. M. Roland, Phys. Rev. Lett. **90**, 175702 (2003)

[6] M. Paluch, C. M. Roland, S. Pawlus, J. Zioło, and K. L. Ngai, Phys. Rev. Lett. **91**, 115701 (2003)

[7] L. Hong, B. Begen, A. Kisliuk, S. Pawlus, M. Paluch, and A. P. Sokolov, Phys. Rev. Lett. **102**, 145502 (2009)

[8] Z. Wojnarowska, C. M. Roland, A. Swiety-Pospiech, K. Grzybowska, and M. Paluch, Phys. Rev. Lett. **108**, 015701 (2012)

[9] M Paluch, K Grzybowska and A Grzybowski, J. Phys.: Condens. Matter **19**, 205117 (2007)

[10] M. Paluch, S. Hensel-Bielowka, and T. Psurek, J. Chem. Phys. **113**, 4374 (2000)

[11] M. Paluch; C. M. Roland; A. Best, J. Chem. Phys. **117**, 1188 (2002)

[12] M. Mierzwa, S. Pawlus, M. Paluch, E. Kaminska, and K. L. Ngai, J. Chem. Phys. **128**, 044512 (2008)

[13] M. Paluch, S. Haracz, A. Grzybowski, M. Mierzwa, J. Pionteck, A. Rivera-Calzada, and C. Leon, J. Phys. Chem. Lett. **1**, 987 (2010)

[14] G. Williams, Trans. Faraday Soc. **60**,1548 (1964)

[15] M. Paluch, J. Gapinski, A. Patkowski, and E. W. Fischer, J. Chem. Phys. **114**, 8048 (2001)

[16] A. Tölle, Rep. Prog. Phys. **64**, 1473 (2001).

[17] C. Dreyfus, A. Aouadi, J. Gapinski, M. Matos-Lopes, W. Steffen, and A. Patkowski, R. M. Pick, Phys. Rev. E **68**, 011204 (2003).

[18] C. Dreyfus, A. Le Grand, J. Gapinski, W. Steffen, and A. Patkowski, Eur. Phys. J. B **42**, 309 (2004)

[19] C. Alba-Simionesco, A. Cailliaux, A. Alegria, and G. Tarjus, Europhys. Lett. **68**, 58 (2004).

[20] R. Casalini and C. M. Roland, Phys. Rev. E **69**, 062501 (2004).

[21] S. Pawlus, R. Casalini, C. M. Roland, M. Paluch, S. J. Rzoska, and J. Ziolo, Phys. Rev. E **70**, 061501 (2004).

[22] R. Casalini and C. M. Roland, Phys. Rev. B **71**, 014210 (2005).

[23] A. Reiser, G. Kasper, and S. Hunklinger, Phys. Rev. B **72**, 094204 (2005); ibid. **74**, 019902(E) (2006).

[24] C. M. Roland, S. Bair, and R. Casalini, J. Chem. Phys. **125**, 124508 (2006).

[25] C. Alba-Simionesco and G. Tarjus, J. Non-Cryst. Solids **352**, 4888 (2006).

[26] U. R. Pedersen, N. P. Bailey, T. B. Schrøder, and J. C. Dyre, Phys. Rev. Lett. **100**, 015701 (2008).

[27] N. P. Bailey, U. R. Pedersen, N. Gnan, T. B. Schrøder, and J. C. Dyre, J. Chem. Phys. **129**, 184507 (2008).

[28] N. P. Bailey, U. R. Pedersen, N. Gnan, T. B. Schrøder, and J. C. Dyre, J. Chem. Phys. **129**, 184508 (2008).

[29] D. Coslovich and C. M. Roland, J. Phys. Chem. B **112**, 1329 (2008).

[30] D. Coslovich and C. M. Roland, J. Chem. Phys. **130**, 014508 (2009).

[31] T. B. Schrøder, U. R. Pedersen, N. P. Bailey, S. Toxvaerd, and J. C. Dyre, Phys. Rev. E **80**, 041502 (2009).

[32] W. G. Hoover and M. Ross, Contemp. Phys. **12**, 339 (1971).

[33] U. R. Pedersen, T. B. Schrøder, and J. C. Dyre, Phys. Rev. Lett. **105**, 157801 (2010).

[34] W. Kob and H. C. Andersen, Phys. Rev. Lett. 73, 1376 (1994).

[35] C. M. Roland, J. L. Feldman, and R. Casalini, J. Non-Cryst. Solids **352**, 4895 (2006)

[36] A. Grzybowski, M. Paluch, and K. Grzybowska, J. Phys. Chem. B **113**, 7419 (2009).

[37] A. Grzybowski, M. Paluch, and K. Grzybowska, Phys. Rev. E **82**, 013501 (2010).

[38] A. Grzybowski, S. Haracz, M. Paluch, and K. Grzybowska, J. Phys. Chem. B **114**, 11544 (2010).

[39] A. Grzybowski, K. Grzybowska, M. Paluch, A. Swiety, and K. Koperwas, Phys. Rev. E **83**, 041505 (2011)

[40] A. Grzybowski, K. Koperwas, and M. Paluch, Phys. Rev. E (in press)

[41] L. Bøhling, T. S. Ingebrigtsen, A. Grzybowski, M. Paluch, J. C. Dyre, Thomas B. Schrøder, arXiv:1112.1602v1 [cond-mat.soft]

[42] T. S. Ingebrigtsen, L. Bøhling, T. B. Schrøder, and J. C. Dyre, J. Chem. Phys. **136**, 061102 (2012)

[43] R. Casalini, U. Mohanty, and C. M. Roland, J. Chem. Phys. **125**, 014505 (2006).

[44] I. Avramov, J. Non-Cryst. Solids **262**, 258 (2000).

[45] A. Grzybowski, M. Paluch, K. Grzybowska, and S. Haracz, J. Chem. Phys. **133**, 161101 (2010)

[46] S. Hensel-Bielowka, J. Ziolo, M. Paluch, and C. M. Roland, J. Chem. Phys. **117**, 2317 (2002)

[47] M. Paluch, C. M. Roland, R. Casalini, G. Meier, A. Patkowski, J. Chem. Phys. **118**, 4578 (2003)

[48] M. Paluch, R. Casalini, and C. M. Roland, Phys. Rev. B **66**, 092202 (2002)

[49] M. Paluch, R. Casalini, A. Best, and A. Patkowski, J. Chem. Phys. **117**, 7624 (2002)

[50] Roland, C. M.; Casalini, R. *Macromolecules* **2003**, *36*, 1361 - 1367.

[51] J. K. McKinney and R. Simha, Macromolecules **7**, 894 (1974).

[52] Z. Wojnarowska, M. Paluch, A. Grzybowski, K. Adrjanowicz, K. Grzybowska, K. Kaminski, P. Wlodarczyk, J. Pionteck, J. Chem. Phys. **131**, 104505 (2009)

[53] We use herein the following values of the parameters of Eq. (2) for VH: $A_1=(0.8666\pm0.0001)\text{cm}^3/\text{g}$, $A_2=(4.90\pm0.02)\cdot10^{-4}\text{cm}^3/(\text{gK})$, $A_3=(4.02\pm0.14)\cdot10^{-7}\text{cm}^3/(\text{gK}^2)$, $B_{T_0}(p_0)=(2805\pm9)$ MPa, $b_2=(4.93\pm0.04)\cdot10^{-3}\text{K}^{-1}$, $\gamma_{EOS}=11.79\pm0.07$, which are found by us from fitting isobaric PVT data reported in Ref. 52, on the assumption that $T_0=319.3$K and $p_0=0.1$MPa in the reference state.

[54] R. Casalini and C. M. Roland, J. Chem. Phys. 119, 11951 (2003)

[55] K. Grzybowska, S. Pawlus, M. Mierzwa, M. Paluch, and K. L. Ngai, J. Chem. Phys. 125, 144507 (2006)

[56] T. Sun, A. S. Teja, J. Chem. Eng. Data **49**, 1311 (2004)

[57] A. Grzybowski; K. Grzybowska; J. Ziolo, and M. Paluch, Phys. Rev. E **74**, 041503 (2006).

[58] A. Grzybowski, K. Kolodziejczyk, K. Koperwas, K. Grzybowska, and M. Paluch, Phys. Rev. B **85**, 220201(R) (2012)

[59] We use herein the following values of the parameters of Eq. (2) for DPG: $A_1=(0.9200\pm0.0003)\text{cm}^3/\text{g}$, $A_2=(4.26\pm0.03)\cdot10^{-4}\text{cm}^3/(\text{gK})$, $A_3=(1.45\pm0.01)\cdot10^{-6}\text{cm}^3/(\text{gK}^2)$, $B_{T_0}(p_0)=(3696\pm11)$ MPa, $b_2=(6.27\pm0.02)\cdot10^{-3}\text{K}^{-1}$, $\gamma_{EOS}=10.23\pm0.03$, which are found by us from fitting the combined and adjusted isobaric PVT data measured at ambient [56] and high [54] pressures, on the assumption that $T_0=195.15$K and $p_0=0.1$MPa in the reference state.

[60] N. Gnan, T. B. Schrøder, U. R. Pedersen, N. P. Bailey, and J. C. Dyre, J. Chem. Phys. **131**, 234504 (2009)

[61] R. Casalini and C.M. Roland, J. Non-Cryst. Solids **353**, 3936 (2007)

[62] A. N. Papathanassiou and I. Sakellis, J. Chem. Phys. **132**, 154503 (2010)

[63] A. N. Papathanassiou, Phys. Rev. E **79**, 032501 (2009)